\documentclass[prd,preprint,nofootinbib]{revtex4}

\usepackage{graphicx}
\usepackage{amssymb}
\usepackage{amsmath}

\begin{document}

\renewcommand{\thefootnote}{\#\arabic{footnote}}
\newcommand{\rem}[1]{{\bf [#1]}}
\newcommand{\gsim}{ \mathop{}_ {\textstyle \sim}^{\textstyle >} }
\newcommand{\lsim}{ \mathop{}_ {\textstyle \sim}^{\textstyle <} }
\newcommand{\vev}[1]{ \left\langle {#1}  \right\rangle }
\newcommand{\bear}{\begin{array}}  
\newcommand {\eear}{\end{array}}
\newcommand{\bea}{\begin{eqnarray}}   
\newcommand{\eea}{\end{eqnarray}}
\newcommand{\beq}{\begin{equation}}   
\newcommand{\eeq}{\end{equation}}
\newcommand{\bef}{\begin{figure}}  
\newcommand {\eef}{\end{figure}}
\newcommand{\bec}{\begin{center}} 
\newcommand {\eec}{\end{center}}
\newcommand{\non}{\nonumber}  
\newcommand {\eqn}[1]{\beq {#1}\eeq}
\newcommand{\la}{\left\langle}  
\newcommand{\ra}{\right\rangle}
\newcommand{\ds}{\displaystyle}

\def\SEC#1{Sec.~\ref{#1}}
\def\FIG#1{Fig.~\ref{#1}}
\def\EQ#1{Eq.~(\ref{#1})}
\def\EQS#1{Eqs.~(\ref{#1})}
\def\lrf#1#2{ \left(\frac{#1}{#2}\right)}
\def\lrfp#1#2#3{ \left(\frac{#1}{#2} \right)^{#3}}
\def\GEV#1{10^{#1}{\rm\,GeV}}
\def\MEV#1{10^{#1}{\rm\,MeV}}
\def\KEV#1{10^{#1}{\rm\,keV}}

\def\lrf#1#2{ \left(\frac{#1}{#2}\right)}
\def\lrfp#1#2#3{ \left(\frac{#1}{#2} \right)^{#3}}
\def \fnl{f_{\rm NL}}

\begin{flushright}
\end{flushright}

\title{
Non-Gaussianity from Baryon Asymmetry
}

\author{Masahiro Kawasaki$^{(a, b)}$, Kazunori Nakayama$^{(a)}$ and Fuminobu Takahashi$^{(b)}$}

\affiliation{%
$^a$ Institute for Cosmic Ray Research,
     University of Tokyo, Chiba 277-8582, Japan\\
$^b$ Institute for the Physics and Mathematics of the Universe, 
     University of Tokyo, Chiba 277-8568, Japan
}

\date{\today}

\begin{abstract}
We study a scenario that large non-Gaussianity arises from the baryon
asymmetry of the Universe. There are baryogenesis scenarios containing
a light scalar field, which may result in baryonic isocurvature
perturbations with some amount of non-Gaussianity.  As an explicit
example we consider the Affleck-Dine mechanism and show that a flat
direction of the supersymmeteric standard model can generate large
non-Gaussianity in the curvature perturbations, satisfying the
observational constraints on the baryonic isocurvature
perturbations. The sign of a non-linearity parameter, $f_{\rm NL}$, is
negative, if the Affleck-Dine mechanism accounts for the observed
baryon asymmetry; otherwise it can be either positive or negative.
\end{abstract}

\preprint{IPMU-08-0061}
\preprint{ICRR-Report-528}

\maketitle

\section{Introduction}
\label{sec:1}

The WMAP results~\cite{Komatsu:2008hk} provided strong support for the
inflation; the observed primordial fluctuations are consistent with
nearly scale-invariant, adiabatic and Gaussian density perturbations,
as predicted by a simple class of inflation models. Those predictions
are derived based on a simple but crude assumption that it is only the
inflaton that acquires sizable quantum fluctuations during inflation.
Its apparent success, however, does not necessarily mean that such a
non-trivial condition is commonly met in the landscape of the
inflation theory.

It is perhaps natural to expect that there are many scalar fields in
Nature. If some of them are light during inflation, they acquire
quantum fluctuations, which may leave slight deviation from the above
properties in e.g. the cosmic microwave background (CMB) anisotropy,
such as the isocurvature perturbations and/or sizable non-Gaussianity.
Interestingly, it was recently reported that large non-Gaussianity was
detected by the analysis on the WMAP 3yr data \cite{Yadav:2007yy}.
The latest WMAP 5yr data seem to have the same tendency, although the
vanishing non-Gaussianity is allowed within 95\%
C.L. \cite{Komatsu:2008hk}.  Those hints on non-Gaussianity may be
originated from such additional light scalars.
 
The non-Gaussian features in the observed CMB can arise from either
adiabatic or isocurvature density perturbations. While the former was
extensively studied in e.g. the curvaton
\cite{Lyth:2001nq,Lyth:2002my}/ungaussiton \cite{Suyama:2008nt}
scenarios, much less attention was paid to the latter case since this
possibility was noted in Refs.~\cite{Linde:1996gt,Boubekeur:2005fj}.

Recently Sekiguchi, Suyama and the current authors systematically
studied how the non-Gaussian isocurvature perturbations exhibit
themselves in the CMB temperature fluctuations \cite{Kawasaki:2008sn}.
In particular, it turned out that the non-Gaussianity in the
isocurvature perturbations is enhanced at large scales, which may be
confirmed (or refuted) by the current and future observations.  In
Ref.~\cite{Kawasaki:2008sn} we also studied the QCD axion as an
example, which generates isocurvature perturbations in the cold dark
matter (CDM).  In this paper we focus a scenario that baryonic
isocurvature perturbations possess non-Gaussian properties.

One of the promising candidates for providing a theory beyond the
standard model (SM) is supersymmetry (SUSY), and the supersymmetric SM
(SSM) contains many flat directions. Those flat directions can play
important roles in cosmology; they generate the baryon asymmetry of
the Universe via the Affleck-Dine (AD) mechanism
\cite{Affleck:1984fy,Dine:1995kz}, and may account for dark matter by
deforming into $Q$-balls
\cite{Coleman:1985ki,Kusenko:1997zq,Dvali:1997qv,Enqvist:1997si}.  If
a flat direction remains light during inflation, it acquires quantum
fluctuations, leading to the baryonic isocurvature fluctuations
\cite{Linde:1985gh,Enqvist:1998pf,Kasuya:2008xp}.  Moreover, as shown in
\cite{Kasuya:2008xp}, the phase direction of the flat direction
generically remains flat in most inflation models in supergravity.
Thus, we expect that the baryonic isocurvature perturbations as well
as the associated non-Gaussianity are generically present in the AD
mechanism.

In this paper we study the non-Gaussian property of the baryonic
isocurvature perturbations produced in the AD mechanism.  We find that
the resultant non-Gaussianity has distinctive features, compared to
those produced in the curvaton/ungaussiton mechanism.  In terms of a
non-linerity parameter, $f_{\rm NL}$, to be defined later, the AD
mechanism predicts a negative value of $\fnl$ if the mechanism
accounts for the observed baryon asymmetry of the Universe; otherwise,
$f_{\rm NL}$ becomes positive or negative, depending on the sign of
baryon asymmetry created by the AD mechanism.  
We would like to emphasize that the AD mechanism can generate a significant amount of
non-Gaussianity while accounting for the {\it total} baryon asymmetry
of the Universe.

This paper is organized as follows.  In Sec.~\ref{sec:NGiso} we
briefly summarize the calculation of non-Gaussianity from isocurvature
perturbations.  Then we discuss non-Gausianity generated by the AD
mechanism in Sec.~\ref{sec:NGAD}.  Sec.~\ref{sec:conclusion} is
devoted to discussion and conclusions.

\section{Non-Gaussianity from isocurvature perturbations} 
\label{sec:NGiso}

We write the spacetime metric as
\begin{equation}
	ds^2\;=\;-{\mathcal N}^2 dt^2 +a^2(t)e^{2\psi} \gamma_{ij} \left ( dx^i + \beta^i dt \right )
	\left ( dx^j + \beta^j dt \right ),
\end{equation}
where ${\mathcal N}$ is the lapse function, $\beta_i$ the shift
vector, $\gamma_{ij}$ the spatial metric, $a(t)$ the background scale
factor, and $\psi$ the curvature perturbation.  We denote by $\zeta$
the curvature perturbation $\psi$ evaluated on the uniform-density
slicing.  The power spectrum $P_\zeta(k)$ and the bispectrum
$B_\zeta(k_1,k_2,k_3)$ of $\zeta$ are defined by the two-point and three-point 
correlation functions as
\bea
	\langle \zeta_{{\vec k_1}} \zeta_{{\vec k_2}} \rangle &\equiv&
	{(2\pi)}^3 	\delta ({\vec k_1}+{\vec k_2}) \,P_\zeta(k_1), \\
	\langle \zeta_{ {\vec k_1}} \zeta_{ {\vec k_2}} \zeta_{ {\vec k_3}} \rangle &\equiv& 
	{(2\pi)}^3 	\delta ({\vec k_1}+{\vec k_2}+{\vec k_3})\, B_\zeta(k_1,k_2,k_3),
\eea
where $\zeta_{\vec k_i}$ is a Fourier component of $\zeta$, i.e.,
$\zeta_{\vec k_i} \equiv \int d^3 x e^{-i {\vec k_i} \cdot {\vec x}}
\zeta ({\vec x})$, and $k_i \equiv |{\vec k_i}|$ for $i = 1,2,3$.  The
non-linearity parameter $f_{\rm NL}$ is defined by~\footnote{
In this paper we consider only the local type non-Gaussianity \cite{Babich:2004gb}.
}
\begin{equation}
	B_\zeta(k_1,k_2,k_3) \;\equiv\; \frac{6}{5}f_{\rm NL} 
	[ P_{\zeta}(k_1) P_{\zeta}(k_2)+2~{\rm perms.} ]. \label{defnl}
\end{equation}

Let us now define the CDM and baryon isocurvature perturbations in the
radiation-dominated universe as
\bea
S_{c \gamma} &\equiv& 3(\zeta_c-\zeta_r),\\
S_{b \gamma} &\equiv& 3(\zeta_b-\zeta_r),
\eea
where $\zeta_{x}$ is the curvature perturbation on a slicing where the
energy density of the component $x$ is spatially uniform, and $x =
\{c,b,r\}$ corresponds to CDM, baryon, and radiation, respectively.
Since the baryonic isocurvature perturbation cannot be distinguished
from the CDM isocurvature one, it is useful to define the effective
CDM isocurvature perturbation as
\begin{equation}
S \;\equiv\;	S_{c\gamma} + R\, S_{b\gamma},
	\label{effS}
\end{equation}
where $R = \Omega_b/\Omega_{\rm CDM} \simeq 0.2$.

We can similarly define the power spectrum $P_S(k)$ and the bispectrum
$B_S(k_1,k_2,k_3)$ of the effective CDM isocurvature perturbation $S$
as
\bea 
\langle S_{{\vec k_1}} S_{{\vec k_2}} \rangle &\equiv& {(2\pi)}^3
\delta ({\vec k_1}+{\vec k_2}) \,P_S(k_1), \label{2S} \\ \langle
S_{{\vec k_1}} S_{{\vec k_2}} S_{{\vec k_3}} \rangle &\equiv&
{(2\pi)}^3 \delta ({\vec k_1}+{\vec k_2}+{\vec k_3})
\,B_S(k_1,k_2,k_3), \label{3S} 
\eea
and the non-linearity parameter $f_S$ is defined by
\begin{eqnarray}
	B_S(k_1,k_2,k_3) \;\equiv\; f_S [ P_{S}(k_1) P_{S}(k_2)+2~{\rm perms.} ]. \label{fS}
\end{eqnarray}

Suppose that the isocurvature perturbation $S$ is sourced by a 
scalar field, $\phi$.  Then $S$ can be expanded in terms of  the fluctuation of $\phi$ as
\begin{equation}
	S \;=\; S_\phi \delta \phi + \frac{1}{2}S_{\phi \phi}(\delta \phi)^2 + \cdots,
	\label{eq:S-expansion}
\end{equation}
where the fluctuation $\delta \phi$ is evaluated when the corresponding scale leaves
the horizon during inflation.  We define the power spectrum of the
scalar field as
\begin{equation}
	\langle \delta \phi_{{\vec k_1}} \delta \phi_{{\vec k_2}} \rangle \;\equiv\;
	{(2\pi)}^3 	\delta ({\vec k_1}+{\vec k_2})P_{\delta \phi}(k_1).
\end{equation}
If the mass of $\phi$ is much smaller than $H_{\rm inf}$, the power
spectrum is approximately given by
\beq
P_{\delta \phi}(k) \;\simeq\; \frac{H_{\rm inf}^2}{2k^3},
\eeq
where $H_{\rm inf}$ is the Hubble parameter during inflation, and we
neglect the tilt of $H_{\rm inf}$ for simplicity.  For later use we
also define the following:
\beq
\Delta_{\delta \phi}^2 \;\equiv \; \frac{k^3}{2 \pi^2} P_{\delta \phi}(k) 
\simeq \lrfp{H_{\rm inf}}{2 \pi}{2}.
\eeq

We can express $P_S$ in terms of the $\delta \phi$ by substituting 
Eq.~(\ref{eq:S-expansion}) into Eq.~(\ref{2S}),
\bea
	P_S(k) &\simeq& \left[  S_\phi^2 + S_{\phi \phi}^2\Delta_{\delta \phi}^2 \ln (kL) \right] 
	P_{\delta \phi}(k), 
\eea
where we have introduced an infrared cutoff $L$, which is set to be of
order of the present Hubble
scale~\cite{Lyth:1991ub,Boubekeur:2005fj,Lyth:2007jh}.  In a similar
way, we can express $B_S$ in terms of $\delta \phi$. The expression
becomes simple when we take the so-called squeezed configuration in
which one of the three wavenumbers is much smaller than the other two
(e.g. $k_1 \ll k_2,k_3$), and it is given by
\bea	
	B_S(k_1,k_2,k_3) &\simeq&  \left[  S_\phi^2 S_{\phi \phi} + 
	S_{\phi \phi}^3 \Delta_{\delta \phi}^2  \ln (k_b L) \right] 
	\left[ P_{\delta \phi}(k_1)P_{\delta \phi}(k_2) + ({\rm 2~perms.})  \right] 
\eea
where $k_b \equiv {\rm min}\{k_1,k_2,k_3\}$.  Thus, for the squeezed
configuration: $k_1 \ll k_2 , k_3$, $f_S$ is given by
\cite{Kawasaki:2008sn}
\begin{equation}
	f_S \;\simeq\; \frac{S_{\phi \phi}}
	{S_\phi^2 + S_{\phi \phi}^2 |\Delta_{\delta \phi}^2| \ln(k_2 L)}.  \label{fSSphi}
\end{equation}
In the following, we take the configuration, $k_1 \ll k_2 , k_3$, as the squeezed configuration.

Let us now relate $f_S$ to $\fnl$.  The curvature perturbation in the
matter dominated era is given by
\begin{equation}
	\zeta = \zeta^{(\rm p)}+\frac{1}{3}S,  \label{zetaS}
\end{equation}
where $\zeta^{(\rm p)}$ denotes the primordial curvature perturbation
created by the inflaton.  We assume that the power spectrum of $\zeta$
is predominantly produced by $\zeta^{(\rm p)}$, while the three-point
correlation function originates from $S$, i.e., $B_\zeta \simeq
B_S/27$.  For the squeezed configuration,  $k_1 \ll k_2 , k_3$,  we obtain
\bea
	f_{\rm NL}^{\rm (iso)} &\simeq&\frac{5}{162} \lrfp{P_S}{P_\zeta}{2} f_S,\non\\
		&\simeq& \frac{5}{162} \lrfp{\Delta_{\delta \phi}^2}{\Delta_\zeta^2}{2} 
		\left(S_\phi^2 + S_{\phi \phi}^2 |\Delta_{\delta \phi}^2| \ln(k_1 L)\right) S_{\phi \phi},
	 \label{fS-fNL}
\eea
where we have used Eq.~(\ref{fSSphi}), and we have defined $\Delta_{\zeta}^2 = k^3 P_\zeta/(2 \pi^2)
\simeq 2.4 \times 10^{-9}$~\cite{Komatsu:2008hk}. 
Here we have written the non-linearity parameter as $f_{\rm NL}^{\rm (iso)}$ 
in order to emphasize that the non-Gaussianity comes from the isocurvature perturbation.
This relation is insensitive to the wavenumbers, 
up to the tilt of the $P_\zeta$ and $P_S$. Note that the sign of $\fnl^{\rm (iso)}$ is determined by
that of $f_S$, or equivalently, $S_{\phi \phi}$.

Lastly we comment on the magnitude of $\fnl$. The observed CMB
temperature fluctuations are consistent with the pure adiabatic
perturbations, and there is a tight constraint on the isocurvature
perturbations. According to the latest WMAP 5yr data~\cite{Komatsu:2008hk}, 
the constraint reads $P_S/P_\zeta \lsim 0.19$ at $95\%$ C.L.
for uncorrelated isocurvature perturbations. Therefore, in order to
have large non-Gaussianity, $|\fnl| \gsim 1$, one can see from
Eq.~(\ref{fS-fNL}) that $|f_S|$ must be at least larger than $3 \times
10^3$.  It should be noticed however that the $f_{\rm NL}$ in
Eq.~(\ref{fS-fNL}) affects the CMB temperature fluctuations in a
completely different way from the conventional $\fnl$ defined for the
adiabatic perturbation.  That is to say, the currently available
constraint on $\fnl$, $-9 < f_{\rm NL} < 111$ at 95\%
C.L. \cite{Komatsu:2008hk}, cannot be applied to the $\fnl^{\rm (iso)}$ in our
case. This is because the constraint is derived assuming that the
non-Gaussianity arises from the adiabatic perturbations. What is more
relevant to the CMB observations is $f_{\rm NL}^{\Delta T}$, an
effective non-linearity parameter defined by the three-point
correlation function of the CMB temperature fluctuations.  For 
isocurvature perturbations with non-Gaussianity, 
$f_{\rm NL}^{\Delta T}$ sensitively depends on the scales of
interest. Indeed, as pointed out in Ref.~\cite{Kawasaki:2008sn},
$f_{\rm NL}^{\Delta T}$ is greatly enhanced as $f_{\rm NL}^{\Delta T}
\sim 100f_{\rm NL}^{\rm (iso)}$ at large scales. No observational
constraint on $f_{\rm NL}^{\Delta T}$ is known yet, and so, we estimate
$f_{\rm NL}^{\rm (iso)}$ instead of $f_{\rm NL}^{\Delta T}$, and take $|f_{\rm
  NL}^{\rm (iso)}| > 1$ as the criterion for ``large" non-Gaussianity. The reader
should keep in mind that the 
$\fnl ^{\rm (iso)}$ affects the CMB temperature fluctuations differently from that defined for
the adiabatic perturbations.
Also note that although Eq.~(\ref{zetaS}) holds only for large scales in the matter dominated era,
$f_{\rm NL}^{\rm (iso)}$ correctly characterizes small scale perturbations in the CMB anisotropy,
once the transfer function of the isocurvature perturbation is taken into account,
as explicitly shown in Ref.~\cite{Kawasaki:2008sn}.

\section{Non-Gaussianity from Affleck-Dine mechanism} \label{sec:NGAD}

There are several baryogenesis mechanisms containing a light scalar.
Among them, we focus on the AD mechanism, which is described briefly below.
We will comment on other mechanisms in Sec.~\ref{sec:conclusion}.

The SSM contains many flat directions consisting of squark, slepton and Higgs fields.
The flat directions are parameterized by composite gauge-invariant
monomial operators such as $udd$ or $LH_u$, and the dynamics of a flat
direction can be expressed in terms of a complex scalar field $\phi$,
dubbed the AD field.
The flat directions of the minimal SSM are classified in Ref.~\cite{Gherghetta:1995dv}. 
We assume that $\phi$ has a nonzero baryon number in the following.

A flat direction has a vanishing scalar potential in the SUSY limit
as long as only renormalizable terms in the superpotential are considered, but
it is lifted by a non-renormalizable operator  in the superpotential:
\begin{equation}
	W_{\rm NR} \;=\; \frac{\phi^n}{nM^{n-3}},
	\label{NR}
\end{equation}
where $M$ denotes an effective cutoff scale for the interaction,
and $n$ is an integer: $n= 4,5,6,7$ and $9$, which depends on flat directions. 
In addition to the non-reanormalizable operator, the flat direction is
lifted by the SUSY breaking effects. 

In gravity-mediated SUSY breaking models,
the scalar potential for the AD field $\phi$ is given by
\begin{equation}
	V_S(\phi)=(m_\phi^2 - cH^2)|\phi |^2 +
	\left (a_m m_{3/2} \frac{\phi^n}{nM^{n-3}} + {\rm h.c.} \right )
	+\frac{|\phi |^{2(n-1)}}{M^{2(n-3)}}, \label{VS}
\end{equation}
where $c$ and $a_m$ are numerical constants of order unity, and $a_m$
is set to be real without loss of generality.  Several comments are in
order. $m_\phi$ is a soft SUSY breaking mass, and $H$ is the Hubble
parameter. We have included here the Hubble-induced mass, which arises
from the quartic couplings between the AD field and the inflaton in
the K\"ahler potential. We have assumed that the sign of the mass term
is negative so that the AD field develops a large expectation value
during inflation.  Note that this mass term is present after inflation
until the reheating is completed. The second term in Eq.~(\ref{VS}) is
the baryon-number violating $A$-term, and the last one is due to the
non-renormalizable operator (\ref{NR}).  We have dropped the so-called
Hubble-induced $A$-terms since they are absent in most inflation
models in supergravity \cite{Kasuya:2008xp}.

%

The AD field also feels finite-temperature effects given by
\cite{Allahverdi:2000zd,Anisimov:2000wx}
\begin{equation}
	V_T(\phi)=\sum_{f_k|\phi |<T}c_k f_k^2 T^2 |\phi |^2
	+ a \alpha(T)^2 T^4 \log \left ( \frac{|\phi |^2}{T^2} \right ), \label{VT}
\end{equation}
where $c_k$ is a constant of order unity, $f_k$ collectively denotes
the gauge and Yukawa couplings for the corresponding AD field, and $a$
is a constant of order unity.  The sign of $a$ depends on flat
directions, and it is determined by the two-loop finite temperature
effective potential.  We assume $a$ to be positive in the
following~\footnote{ If $a$ is negative, the AD field may be trapped
  by the negative thermal logarithmic potential. Some explicit examples of the flat directions
  having the negative corrections are given in Ref.~\cite{Kasuya:2003va}.  If the trap lasts long enough,
  the AD mechanism may not work \cite{Kasuya:2003yr}.  }.
Those thermal effects are known to significantly affect the final
baryon asymmetry \cite{Fujii:2001zr}.

Now let us take a closer look at the dynamics of the AD field. During inflation
the AD field stays at the potential minimum,
\begin{equation}
	|\phi | \;=\; \left(HM^{n-3}\right)^{1/(n-2)},
\end{equation}
which is determined by the balance between the
Hubble mass term and the non-renormalizable term.
After inflation, the Hubble parameter decreases with time, and so does
the minimum.  The radial component of $\phi$ continues to track the
minimum until it begins to oscillate. The oscillations start
when the Hubble parameter becomes equal to $H_{\rm os}$, given by
\begin{equation}
	H_{\rm os}^2 \;=\; m_\phi^2 + \sum_{f_k|\phi |<T}c_k f_k^2 T_{\rm os}^2 + a \alpha(T)^2 
	\frac{T_{\rm os}^4}{|\phi |^2}.
\end{equation}
Here the subscript ``os'' denotes that the variable should be
evaluated at the beginning of the oscillations.  The field value at
which the AD field begins to oscillate is given by
$|\phi_{\rm os}| = (H_{\rm os} M^{n-3})^{1/(n-2)}$.
On the other hand, the phase component of the AD field, $\theta \equiv
{\rm arg}[\phi]$, is almost massless during and after inflation, due
to the absence of the Hubble-induced mass term. (Note that 
 the mass along the phase direction arises only from the baryon-number
violating $A$-term.)  Therefore $\theta$ remains at the initial value
set during inflation, until the AD field starts to oscillate. In the
following we refer the initial value by $\theta$.

When the $\phi$ starts to oscillate, it is also kicked into the phase
direction due to the baryon-number violating $A$-term. Most of the
baryon asymmetry is created at this moment. The angular momentum of
the motion in the complex plane of $\phi$ is related to the baryon
number density:
\begin{equation}
	n_{B}^{\rm (AD)}\;=\;i(\dot \phi^* \phi - \phi^* \dot \phi). \label{nB}
\end{equation}
The baryon-to-entropy ratio created by the AD mechanism is estimated as
\begin{equation}
	\frac{n_B^{\rm (AD)}}{s} \;\sim\; \lambda 
	\frac{m_{3/2} |\phi _{\rm os}|^2 T_R}{H_{\rm os}^2 M_P^2} \sin(n\theta),
	\label{nbs}
\end{equation}
where $\lambda = |a_m|(n-2)/3(n-3)$ is a constant of order unity.  Note that
$\theta$ serves as a CP phase for the successful baryogenesis.  Here we
have assumed that the oscillations begin before the reheating completes.  
The AD mechanism can account for the observed baryon asymmetry of
the Universe, $n_B/s \simeq 8.8\times
10^{-11}$ \cite{Komatsu:2008hk},  for appropriate choices of the cutoff scale $M$ and the
reheating temperature $T_R$.

Now let us consider the fluctuations of the AD field 
\cite{Enqvist:1998pf,Kasuya:2008xp,Riotto:2008gs}.  Since the
angular direction of the AD field does not receive sizable
corrections during inflation~\cite{Kasuya:2008xp}, it has an unsuppressed quantum
fluctuation $\delta \theta$. We define the power spectrum of $\delta \theta$ as
\begin{equation}
	\langle \delta \theta_{{\vec k_1}} \delta \theta_{{\vec k_2}} \rangle \;\equiv\;
	{(2\pi)}^3 	\delta ({\vec k_1}+{\vec k_2})P_{\delta \theta}(k_1).
\end{equation}
The magnitude of the fluctuations is given by
\beq
\Delta_{\delta \theta}^2 \;\equiv \; \frac{k^3}{2 \pi^2} P_{\delta \theta}(k) 
\simeq \lrfp{H_{\rm inf}}{2 \pi |\phi_{\rm inf}|}{2}.
\eeq
where $|\phi_{\rm inf}|=(H_{\rm inf} M^{n-3})^{1/(n-2)}$.  This
fluctuation of the AD field results in the baryonic isocurvature
fluctuation through the AD baryogenesis mechanism,
\begin{equation}
	S_{b\gamma} \;=\; \frac{\delta n_B^{(\rm AD)}}{n_B}
	=r\left[ n \cot (n \theta) \delta \theta - \frac{n^2}{2}(\delta \theta)^2  +\cdots \right],
	\label{Sb}
\end{equation}
where $r\equiv n_B^{(\rm AD)}/n_B$ represents the fraction of the baryon
number created by the AD mechanism to the total baryon
number. If the AD mechanism is responsible for the total baryon asymmetry,
$r$ equals to $1$. In the presence of the other baryogenesis, $r$ can take any values in principle,
and it can be even negative. 
As we will see,  the isocurvature perturbation and/or non-Gaussianity
  induced by the AD field can be significant for both cases of $r=1$ and $|r| \ll 1$.
  
\begin{figure}[t]
   \includegraphics[width=0.6\linewidth]{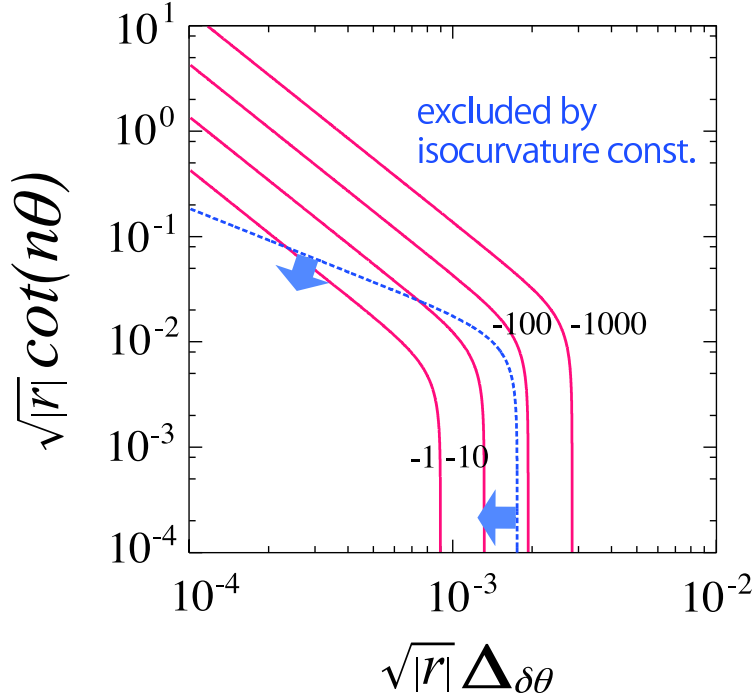} 
   \caption{
   Contour of ${\rm sgn}(r) \fnl^{\rm (iso)} = -1, -10, -100$ and $-1000$ (solid lines), together
   with the constraint on the isocurvature perturbation, $P_S/P_\zeta \lsim 0.19$ (dotted line). 
   The region above the constraint is excluded. 
   We have set $n=6$ and approximated $\ln(kL) \sim 1$. 
   }
   \label{fig:contour}
\end{figure}

The effective CDM isocurvature perturbation $S = RS_{b\gamma}$
is similarly expanded as
\begin{equation}
	S \;=\; S_\theta \delta \theta + \frac{1}{2}S_{\theta \theta} \left ( \delta \theta \right )^2+\cdots,
\end{equation}
with
\bea
S_\theta &\equiv&Rrn\cot(n \theta),\\
S_{\theta \theta}&\equiv&-Rrn^2.
\eea
Eq.~(\ref{fSSphi}) can be rewritten as
\bea
	f_S &\simeq& \frac{S_{\theta \theta}}
	{S_\theta^2 + S_{\theta \theta}^2 |\Delta_{\delta \theta}^2| \ln(k_2L)}, \non\\
		&\simeq&\frac{-1}{R r} \left(\cot^2 (n \theta)+n^2 |\Delta_{\delta \theta}^2|  \ln (k_2L)\right)^{-1}.
	 \label{fSStheta}
\eea
Using \EQ{fS-fNL}, we obtain
\bea
\fnl ^{\rm (iso)}&\simeq& \frac{5}{162} \lrfp{\Delta_{\delta \theta}^2}{\Delta_\zeta^2}{2} 
		\left(S_\theta^2 + S_{\theta \theta}^2 |\Delta_{\delta \theta}^2| \ln(k_1 L)\right) S_{\theta \theta},\non\\
		&=&-\frac{5 R^3 n^4}{162}  r^3 \lrfp{\Delta_{\delta \theta}^2}{\Delta_\zeta^2}{2}  
		 \left(\cot^2 (n \theta)+n^2 |\Delta_{\delta \theta}^2|  \ln (k_1 L)\right).  \label{fNL-Rnr}
\eea
Note that $\fnl^{\rm (iso)}$ is negtive(positive) for a positive(negative) value of $r$. 
The constraint on the isocurvature perturbation, $P_S/P_\zeta \lsim 0.19$ now reads
\beq
(R n r)^2  \left(\cot^2 (n \theta)+n^2 |\Delta_{\delta \theta}^2|  \ln (k L)\right)
\Delta_{\delta \theta}^2 \lesssim 4.6 \times 10^{-10}.
\label{iso}
\eeq
In \FIG{fig:contour} we have plotted the contours of 
${\rm sgn}(r) \fnl^{\rm (iso)} = -1, -10, -100$ and $-1000$,
together with the constraint on the isocurvature perturbations (\ref{iso}). In order to
avoid the constraint, $\cot (n \theta)$ and $\Delta_{\delta \theta}$ are limited to the
following ranges:
\bea
\sqrt{|r|} \cot(n\theta) &\lesssim& O(0.1),\\
\sqrt{|r|} \Delta_{\delta\theta} &\lesssim& O(10^{-3}),
\eea
and the non-Gaussianity parameter $\fnl^{\rm (iso)}$ is bounded as $|\fnl^{\rm (iso)}| \lsim 60$.


\begin{figure}[t]
   \includegraphics[width=0.8\linewidth]{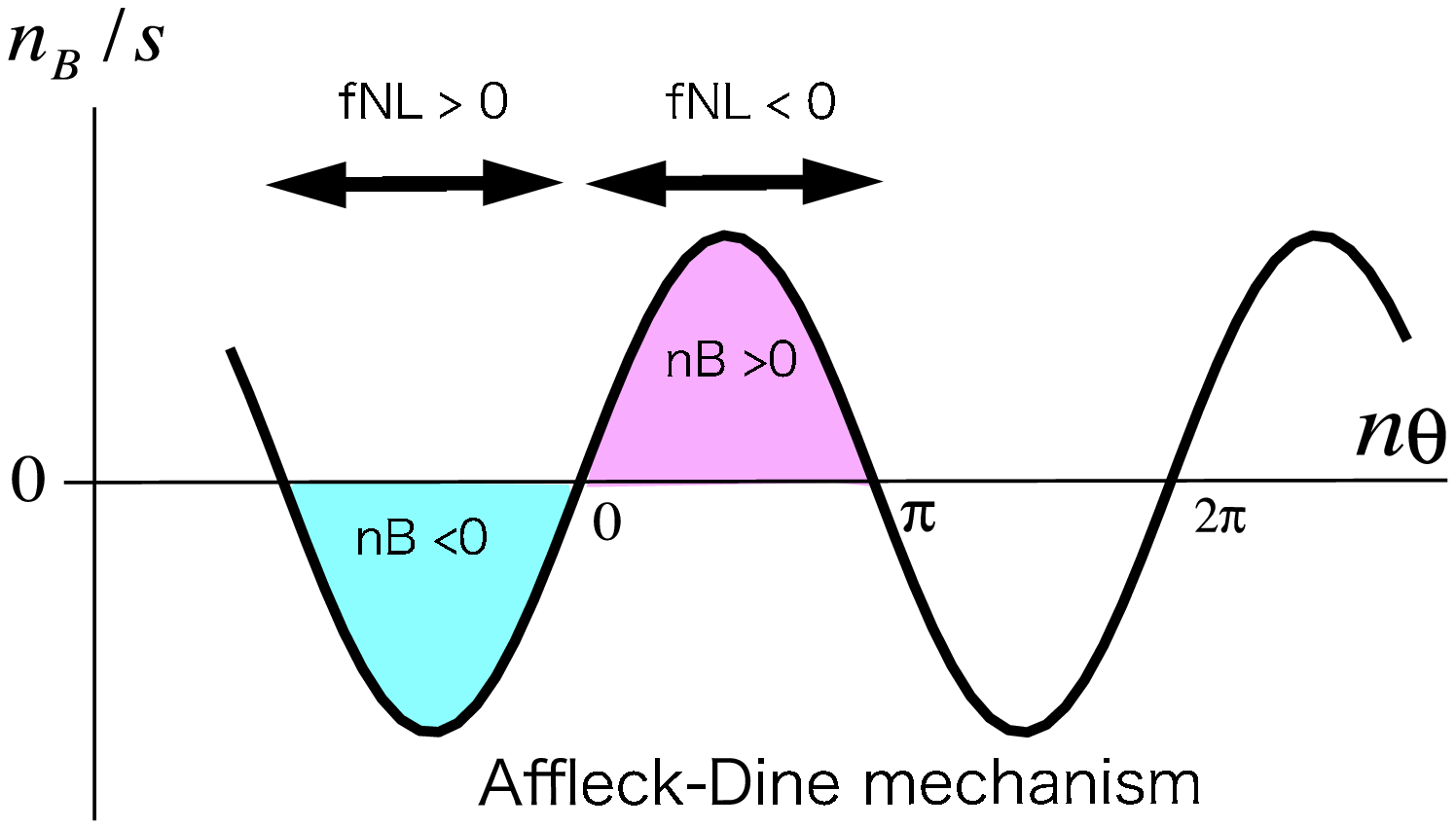} 
   \caption{
   A schematic picture for the baryon asymmetry generated by the AD mechanism.
   We indicate the regions where $f_{\rm NL}^{\rm (iso)}$ is predicted to be positive or negative.
   Roughly speaking, the sign of $\fnl^{\rm (iso)}$ is determined by the curvature of the curves,
   since the coefficient of the second order perturbations is determined by the second
   derivative of the baryon number.
   }
   \label{fig:AD}
\end{figure}

Lastly let us make a comment on a characteristic behavior of the baryonic
isocurvature perturbations (\ref{Sb}).  The second
order term , which represents the non-Gaussian
perturbation, dominates over the linear fluctuation in the limit $\cot (n \theta) \to 0$.  In this
limit the baryon number itself created by the AD mechanism approaches
to the maximum (see (\ref{nbs})). 
In Fig.~\ref{fig:AD} we show schematically the baryon
asymmetry generated by the AD mechanism as a function of $\theta$. 
We have indicated the regions where
$f_{\rm NL}^{\rm (iso)}$ is predicted to be positive or negative.  This helps us 
understand why $f_{\rm NL}^{\rm (iso)}$ becomes negative when the positive baryon
number is generated.


\subsection{Gravity-mediated SUSY breaking models}

In gravity-mediation models, the whole potential of the AD field is
given by Eqs.~(\ref{VS}) and (\ref{VT}).  In the following we assume
$n=6$~\footnote{
In the case of $n=4$, the allowed regions become smaller than those in the
case of $n=6$.
}.  For simplicity, we do not
  consider the case that oscillation due to the thermal mass term
  occurs.  This assumption is justified since we are interested in the
  regime $|\phi| \gg T$ to avoid the gravitino overproduction
  \cite{Moroi:1993mb,Kawasaki:2004yh} and hence particles that couple
  to the AD field cannot be thermalized. 
The AD field begins to oscillate due to thermal logarithmic
term if the reheating temperature $T_R$ satisfies a condition
(hereafter we take $a=1$),
\begin{equation}
\begin{split}
	T_R \;\gtrsim\; T_R^c&\;\equiv\; \frac{1}{\alpha}\left ( \frac{m_\phi^3 M^3}{M_P^2} \right )^{1/4} \\
	& \;\sim \;1.1\times 10^6~{\rm GeV} \left ( \frac{0.1}{\alpha} \right )
	\left ( \frac{m_\phi}{1~{\rm TeV}} \right )^{3/4}
	\left ( \frac{M}{10^{16}~{\rm GeV}} \right )^{3/4}.
\end{split}
\end{equation}
Otherwise it begins its oscillations by the soft mass term.
Thus $r = n_B^{\rm (AD)}/n_B$ is estimated as
\begin{equation}
	r\simeq \left \{
	\begin{array}{ll}
	0.27\displaystyle
	\left ( \frac{m_{3/2}}{1~{\rm TeV}} \right )\left ( \frac{T_R}{10^4~{\rm GeV}} \right )
	\left ( \frac{1~{\rm TeV}}{m_\phi} \right )^{3/2}
	\left ( \frac{M}{10^{16}~{\rm GeV}} \right )^{3/2} \sin (n \theta)
	&~~~{\rm for}~~~T_R < T_R^c \\
	0.35\displaystyle
	\left ( \frac{0.1}{\alpha} \right )^{2}
	\left ( \frac{m_{3/2}}{1~{\rm TeV}} \right ) 
	\left ( \frac{10^8~{\rm GeV}}{T_R} \right )
	\left ( \frac{M}{10^{16}~{\rm GeV}} \right )^{3} \sin (n \theta)
	&~~~{\rm for}~~~T_R > T_R^c 
	\end{array} \right. . \label{rgrav}
\end{equation}
The CDM isocurvature perturbation $S$ is calculated as
\begin{equation}
	S \simeq \left \{
	\begin{array}{ll}
	3\times 10^{-5}r^{1/2}\displaystyle
	\left ( \frac{m_{3/2}}{1~{\rm TeV}} \right )^{1/2}
	\left ( \frac{1~{\rm TeV}}{m_\phi} \right )^{3/4}
	\left ( \frac{T_R}{10^4~{\rm GeV}} \right )^{1/2}
	\displaystyle \left ( \frac{H_{\rm inf}}{10^{14}~{\rm GeV}} \right )^{3/4} 
	\left ( \frac{\cot (n\theta)}{10^{-2}} \right ) \\
	~~~~~\times \sin ^{1/2}(n\theta) ~~~~~{\rm for}~~~T_R < T_R^c \\
	5\times 10^{-5}r^{3/4} \displaystyle
	\left ( \frac{0.1}{\alpha} \right )^{1/2}
	\left ( \frac{m_{3/2}}{1~{\rm TeV}} \right )^{1/4}
	\left ( \frac{10^8~{\rm GeV}}{T_R} \right )^{1/4}
	\displaystyle \left ( \frac{H_{\rm inf}}{10^{14}~{\rm GeV}} \right )^{3/4} 
	\left ( \frac{\cot (n\theta)}{10^{-2}} \right ) \\
	~~~~~\times \sin^{1/4} (n \theta)~~~~~{\rm for}~~~T_R > T_R^c 
	\end{array} \right. , \label{Sgrav}
\end{equation}
if the linear term in Eq.~(\ref{Sb}) dominates.
This is constrained as $\Delta_S \lesssim 2.2\times 10^{-5}$ from WMAP5 results, as stated before.
Then from Eq.~(\ref{fS-fNL}) we can calculate $f_{\rm NL}^{\rm (iso)}$ as
\begin{equation}
	f_{\rm NL}^{\rm (iso)}\simeq \left \{
	\begin{array}{ll}
	-2\times 10^{2}r\displaystyle
	\left ( \frac{m_{3/2}}{1~{\rm TeV}} \right )^{2}
	\left ( \frac{T_R}{10^4~{\rm GeV}} \right )^2
	\left ( \frac{1~{\rm TeV}}{m_\phi} \right )^{3}
	\displaystyle \left ( \frac{H_{\rm inf}}{10^{14}~{\rm GeV}} \right )^{3} 
	\left ( \frac{\cot (n\theta)}{10^{-2}} \right )^2 \\
	~~~~~\times \sin^2 (n \theta) ~~~~~{\rm for}~~~T_R < T_R^c \\
	-1\times 10^{3}r^2 \displaystyle
	\left ( \frac{0.1}{\alpha} \right )^{2}
	\left ( \frac{m_{3/2}}{1~{\rm TeV}} \right )
	\left ( \frac{10^8~{\rm GeV}}{T_R} \right )
	\displaystyle \left ( \frac{H_{\rm inf}}{10^{14}~{\rm GeV}} \right )^{3} 
	\left ( \frac{\cot (n\theta)}{10^{-2}} \right )^2 \\
	~~~~~\times  \sin^{} (n \theta)~~~~~{\rm for}~~~T_R > T_R^c 
	\end{array} \right. . \label{fNLgrav}
\end{equation}
Thus depending on the sign of the CP phase ($\sin (n \theta)$) (or equivalently the
baryon asymmetry), the non-linearity parameter can be either positive or negative.

Here we mention the effect of $Q$-ball formation.  It is known that if a
scalar field has a conserved global $U(1)$ charge~\footnote{
Precisely speaking, the $U(1)$ symmetry is explicitly broken by the $A$-term in the AD 
mechanism. Nevertheless the effect of the breaking
is very small after the AD field starts oscillating~\cite{Kawasaki:2005xc}.
} and if the scalar
potential becomes flatter than a quadratic potential at larger field values, there exists a stable
configuration of the scalar field, called $Q$-ball, whose stability is
ensured by the $U(1)$ symmetry \cite{Coleman:1985ki}.  In the context of
AD mechanism, this conserved $U(1)$ charge is the baryon number. 
In the AD mechanism, the scalar potential tends to be flatter than the quadratic term
due to the renormalization group effects. Thus
$Q$-balls are generically formed in the AD mechanism~\cite{Kusenko:1997zq,Dvali:1997qv,Enqvist:1997si}.

The charge of a $Q$-ball is given by
\begin{equation}
	Q \;\simeq\; \gamma \left ( \frac{\phi_{\rm os}}{m_\phi} \right )^2
	\times \left \{
	\begin{array}{ll}
		\epsilon ~~~&{\rm for}~~~\epsilon > \epsilon_c \\
		\epsilon_c ~~~&{\rm for}~~~\epsilon < \epsilon_c
	\end{array} \right. ,
\end{equation}
where $\gamma \sim 6\times 10^{-3}$ and $\epsilon_c \sim 0.01$ from the lattice calculation
\cite{Kasuya:1999wu}.
The parameter $\epsilon$ is called the ellipticity parameter, defined by
the ratio of the minor and major axes of the orbit of the AD field $\phi$.
It is roughly estimated as
\begin{equation}
	\epsilon 
	\;\sim\;  \frac{m_{3/2}}{H_{\rm os}} \sin(n\theta).
\end{equation}
If $Q \gg 10^{20}$, the decay temperature of the $Q$-ball becomes smaller
than the freeze-out temperature of the lightest supersymmetric particle (LSP)
and hence LSPs emitted by the $Q$-ball decay may be overproduced, depending on the
self-annihilation cross section of the LSP \cite{Enqvist:1997si,Fujii:2001xp,Seto:2005pj,Kawasaki:2007yy}. 
 The charge of the $Q$-ball, $Q$, is estimated as
\begin{equation}
	Q \simeq \left \{
	\begin{array}{ll}
	2\times 10^{17} \displaystyle
	\left ( \frac{m_{3/2}}{m_\phi} \right )
	\left ( \frac{1~{\rm TeV}}{m_{\phi}} \right )^{3/2}
	\left ( \frac{M}{10^{16}~{\rm GeV}} \right )^{3/2}
	~~~{\rm for}~~~T_R < T_R^c \\
	2 \times 10^{11}\displaystyle
	\left ( \frac{0.1}{\alpha} \right )^{2}
	\left ( \frac{10^8~{\rm GeV}}{T_R} \right )^2
	\left ( \frac{M}{10^{16}~{\rm GeV}} \right )^{3}
	~~~{\rm for}~~~T_R > T_R^c 
	\end{array} \right. .
	\label{Qgrav}
\end{equation}
If some fraction of LSP dark matter comes from the $Q$-ball decay,
it produces the additional CDM isocurvature
perturbation with some amount of non-Gaussianity.  Furthermore,
the decay rate of the $Q$-ball depends on its charge $Q$ and hence the decay rate
also becomes the source for the isocurvature fluctuations~\cite{Hamaguchi:2003dc}
as well as non-Gaussianity.  However, in the interesting parameter regions, 
we have checked that the $Q$-balls
evaporate in the high-temperature plasma and hence those effects of 
the $Q$-ball decay are negligible.

\begin{figure}[t]
   \includegraphics[width=0.8\linewidth]{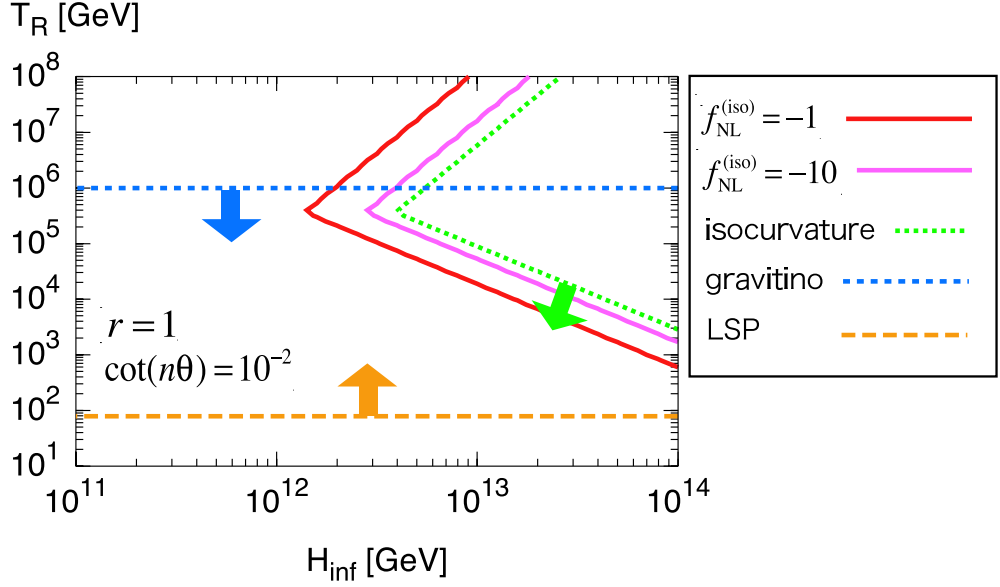}
   \caption{
   Constraints on $(H_{\rm inf},T_R)$ for $m_{3/2}=1$~TeV and
   $\cot (n \theta)=10^{-2}$.
   We have fixed $r=1$.
   The lines show :
   $f_{\rm NL}^{\rm (iso)}=-1$ (red solid), $-10$ (purple solid), isocurvature bound (green dashed), 
   gravitino overproduction bound (blue dotted), 
   LSP overproduction from $Q$-ball decay (orange long-dashed).
   Arrows indicate the allowed region.
   }
   \label{fig:ADm1TeV}
\end{figure}

In Fig.~\ref{fig:ADm1TeV} the contours of $f_{\rm NL}^{\rm (iso)}=-1$ and $-10$ are
shown by the red solid and purple solid lines on the $(H_{\rm inf}, T_R)$
plane for $m_{3/2}=1$~TeV and $\cot (n \theta)=10^{-2}$.  Here we have
fixed $r=1$, i.e., the AD mechanism creates the total baryon asymmetry of the
Universe.  We also show the constraints from the isocurvature perturbation
(green dashed), baryon overproduction (brown dot-dashed), gravitino
overproduction (blue dotted), and LSP overproduction from $Q$-balls
(orange long-dashed). 
To be conservative, we have estimated LSP abundance by neglecting the annihilation of the LSP,
and we set the LSP mass to be 100~GeV.
One can see that a significant amount of
non-Gaussianity can be generated without conflicting the isocurvature constraint
for the Hubble scale during inflation $H_{\rm inf} \gtrsim 10^{12}~$GeV. 
For $\cot (n\theta) \gtrsim
0.1$, the isocurvature constraint becomes more stringent and large
$f_{\rm NL}^{\rm (iso)}$ cannot be generated without conflicting the isocurvature
bound (see also \FIG{fig:contour}).


\begin{figure}[t]
   \includegraphics[width=0.8\linewidth]{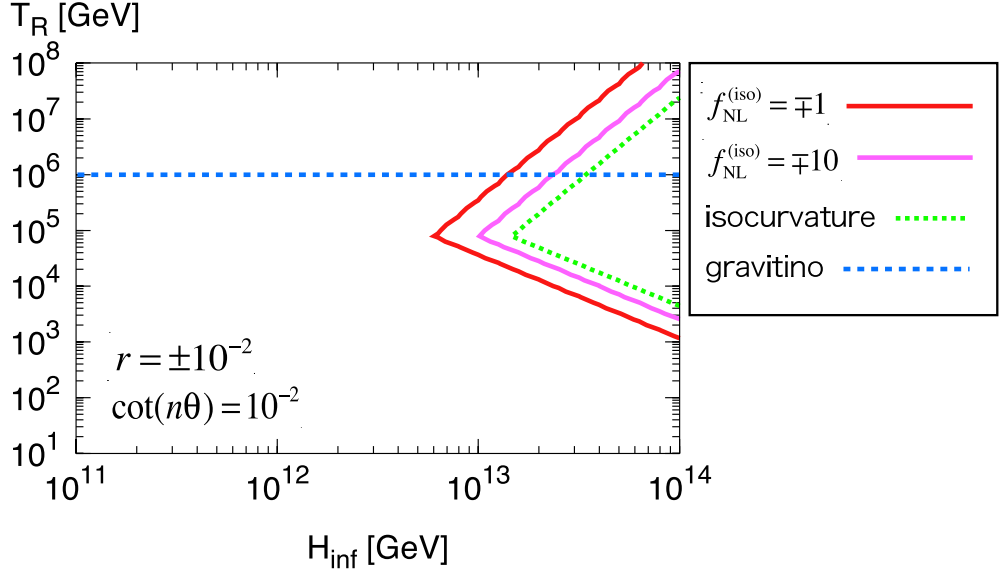}
   \caption{
   	Same as Fig.~\ref{fig:ADm1TeV}, but for $r=\pm 10^{-2}$.
	The sign of $f_{\rm NL}^{\rm (iso)}$ depends on the sign of $r$.
   }
   \label{fig:ADm1TeVr1e-2}
\end{figure}

In the above arguments we have assumed that the AD mechanism provides
the total baryon number of the Universe.  However, the AD mechanism may
create only small fraction of the baryon asymmetry,
most of which is dominantly generated by another baryogenesis
mechanism.  In this case $r$ can be much smaller than unity.  In
Fig.~\ref{fig:ADm1TeVr1e-2} we similarly show the constraints on $(H_{\rm inf},T_R)$
 for $r=\pm 10^{-2}$, $m_{3/2}=1$~TeV and $\cot (n
\theta)=10^{-2}$.  If $r$ is positive (negative), $f_{\rm NL}^{\rm (iso)}$ becomes
negative (positive).  Thus large positive value of $f_{\rm NL}^{\rm (iso)}
(\gtrsim 1)$ can be obtained if the AD mechanism creates small amount
of baryon asymmetry with the negative sign.
In this case the larger inflationary scale ($H_{\rm inf} \gtrsim 10^{13}$~GeV)
than the case of $r=1$ is required (see Eq.~(\ref{fNL-Rnr})).

It may also be useful to rewrite Eqs.~(\ref{Sgrav}) and (\ref{fNLgrav})
by substituting (\ref{rgrav}),
\begin{equation}
	S\simeq \left \{
	\begin{array}{ll}
	2\times 10^{-5}\displaystyle
	\left ( \frac{m_{3/2}}{1~{\rm TeV}} \right )
	\left ( \frac{T_R}{10^4~{\rm GeV}} \right )
	\left ( \frac{1~{\rm TeV}}{m_\phi} \right )^{3/2}
	\left ( \frac{M}{10^{16}~{\rm GeV}} \right )^{3/4}
	\displaystyle \left ( \frac{H_{\rm inf}}{10^{14}~{\rm GeV}} \right )^{3/4}
	\left ( \frac{\cot (n\theta)}{10^{-2}} \right ) \\
	~~~~~\times \sin (n\theta)~~~~~{\rm for}~~~T_R < T_R^c \\
	2\times 10^{-5}\displaystyle
	\left ( \frac{0.1}{\alpha} \right )^{2}
	\left ( \frac{m_{3/2}}{1~{\rm TeV}} \right ) 
	\left ( \frac{10^8~{\rm GeV}}{T_R} \right )
	\left ( \frac{M}{10^{16}~{\rm GeV}} \right )^{9/4}
	\displaystyle \left ( \frac{H_{\rm inf}}{10^{14}~{\rm GeV}} \right )^{3/4}
	\left ( \frac{\cot (n\theta)}{10^{-2}} \right ) \\
	~~~~~\times \sin (n\theta)~~~~~{\rm for}~~~T_R > T_R^c 
	\end{array} \right. ,
\end{equation}
and 
\begin{equation}
	f_{\rm NL}^{\rm (iso)}\simeq \left \{
	\begin{array}{ll}
	-6\times 10^{1}\displaystyle
	\left ( \frac{m_{3/2}}{1~{\rm TeV}} \right )^{3}
	\left ( \frac{T_R}{10^4~{\rm GeV}} \right )^3
	\left ( \frac{1~{\rm TeV}}{m_\phi} \right )^{9/2}
	\left ( \frac{M}{10^{16}~{\rm GeV}} \right )^{3/2}
	\displaystyle \left ( \frac{H_{\rm inf}}{10^{14}~{\rm GeV}} \right )^{3} 
	\left ( \frac{\cot (n\theta)}{10^{-2}} \right )^2 \\
	~~~~~\times \sin^3 (n \theta)~~~~~{\rm for}~~~T_R < T_R^c \\
	-1\times 10^{2}\displaystyle
	\left ( \frac{0.1}{\alpha} \right )^{6}
	\left ( \frac{m_{3/2}}{1~{\rm TeV}} \right )^3 
	\left ( \frac{10^8~{\rm GeV}}{T_R} \right )^3
	\left ( \frac{M}{10^{16}~{\rm GeV}} \right )^{6}
	\displaystyle \left ( \frac{H_{\rm inf}}{10^{14}~{\rm GeV}} \right )^{3} 
	\left ( \frac{\cot (n\theta)}{10^{-2}} \right )^2\\
	~~~~~\times \sin^3 (n \theta)~~~~~{\rm for}~~~T_R > T_R^c 
	\end{array} \right. .
\end{equation}

\begin{figure}[h]
   \includegraphics[width=0.8\linewidth]{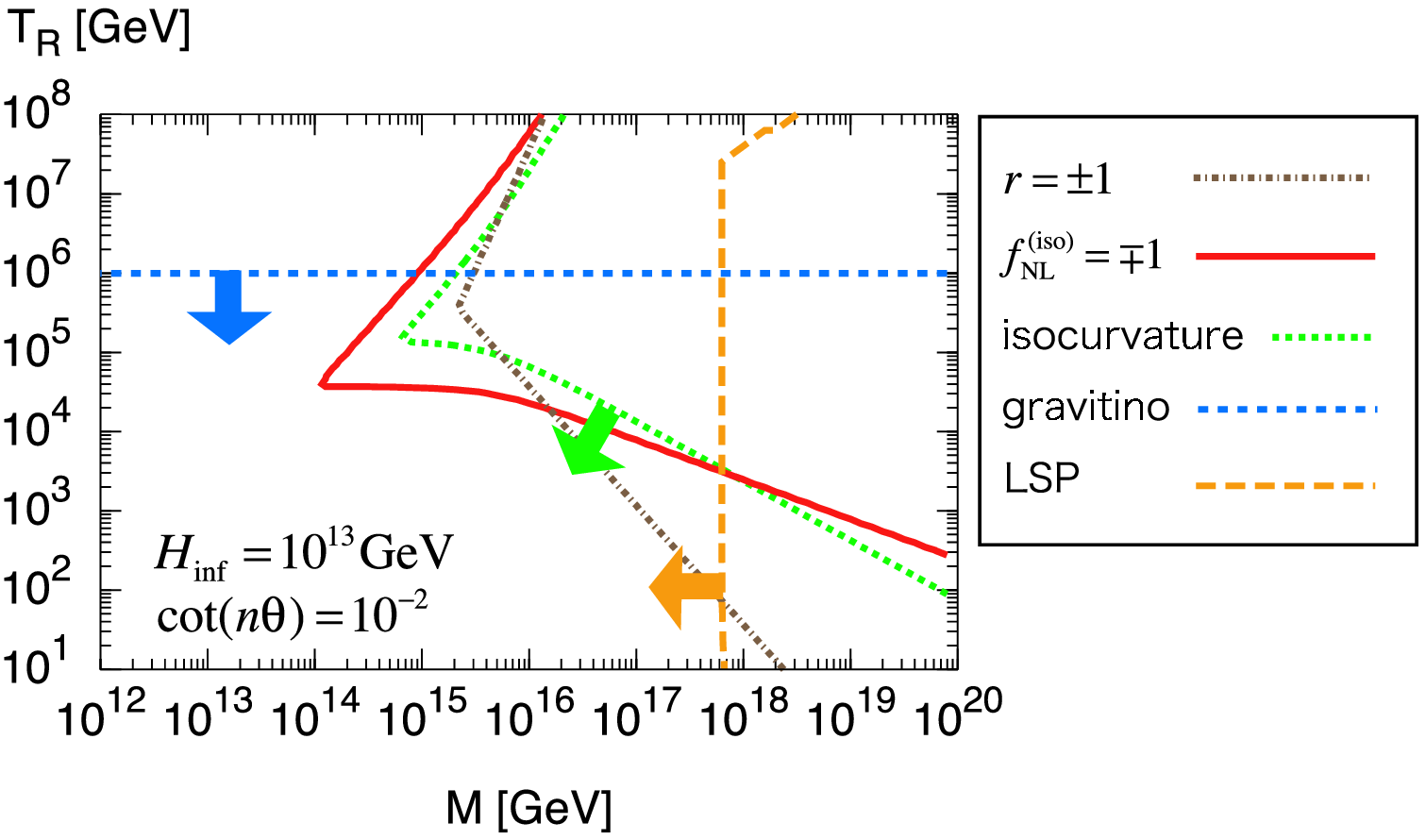}
   \caption{
   Constraints on $(M,T_R)$ for $m_{3/2}=1$~TeV, $H_{\rm inf}=10^{13}~$GeV and
   $\cot (n \theta)=10^{-2}$.
   The lines show :
   $f_{\rm NL}^{\rm (iso)}=-1$ (red solid), isocurvature bound (green dashed), 
   gravitino overproduction bound (blue dotted), baryon overproduction (brown dot-dashed) and
   LSP overproduction from $Q$-ball decay (orange long-dashed).
  Arrows indicate the allowed region.
	   }
   \label{fig:ADm1TeVM}
\end{figure}

We also show the constraints on $(M,T_R)$ plane in Fig.~\ref{fig:ADm1TeVM} without fixing $r$.
Here we take $m_{3/2}=1~$TeV, $\cot (n \theta)=10^{-2}$ and $H_{\rm inf}=10^{13}~$GeV.
The red solid line shows $f_{\rm NL}^{\rm (iso)}=-1$.
One can see that, for $|r| \leq 1$, the cutoff scale $M$ should be smaller than around $10^{16}~$GeV
for generating large non-Gaussianity.


\subsection{Gauge-mediated SUSY breaking models}

In gauge-mediated SUSY breaking models (GMSB) \cite{Giudice:1998bp}, 
messenger fields mediate the SUSY breaking effect to the SSM sector.
Here we assume the direct mediation scenario, that is, 
the SUSY breaking field $S$ couples to messenger fields 
($\Psi, \bar \Psi$) in the superpotential as $W=S\Psi \bar \Psi$.
In this type of models, the gauge mediation is suppressed 
for the scale $|\phi |\gtrsim \langle S \rangle$ 
where $\langle S \rangle$ is the vacuum expectation value of $S$.
Instead there appears a logarithmic correction to the potential as \cite{de Gouvea:1997tn}
\begin{equation}
\begin{split}
	V_S^{(\rm gauge)}(\phi)=&(m_{3/2}^2 - cH^2)|\phi |^2 
	+ V_0\left ( \log \frac{|\phi|^2}{\langle S \rangle^2} \right )^2 \\
	&+\left (a_m m_{3/2} \frac{\phi^n}{nM^{n-3}} + {\rm h.c.} \right )
	+\frac{|\phi |^{2(n-1)}}{M^{2(n-3)}}~~~~~(|\phi| > \langle S \rangle),  \label{Vgauge}
\end{split}
\end{equation}
where $V_0$ is given by
\begin{equation}
	V_0 \equiv M_F^4 = m_\phi^2 \langle S \rangle^2 
	\simeq \left ( \frac{\alpha_i}{4\pi} \right )^2F_S^2,
\end{equation}
where $\alpha_i$ is the gauge couplings relevant for the AD field with
$i$ denoting the gauge groups and $F_S = \sqrt{3}m_{3/2}M_P$ for
vanishing cosmological constant.  Thus the whole potential of the AD
field is the sum of Eq.~(\ref{VT}) and (\ref{Vgauge}).  
The AD field begins to oscillate due to thermal logarithmic term for
\begin{equation}
\begin{split}
	T_R \;\gtrsim\; T_R^c &\equiv \frac{(MV_0)^{3/10}}{\alpha M_P^{1/2}} \\
	& \sim 5.3\times 10^4~{\rm GeV}
	\left ( \frac{0.1}{\alpha} \right )
	\left ( \frac{m_{3/2}}{1~{\rm MeV}} \right )^{3/5}
	\left ( \frac{M}{10^{16}~{\rm GeV}} \right )^{3/10}.
\end{split}
\end{equation}
For $T_R < T_R^c$, the AD field begins to oscillate due to the
logarithmic potential coming from the gauge mediation
effects.\footnote{ We found that in the interesting parameter regions
  in the following, the oscillation due to the $m_{3/2}^2|\phi|^2$
  term does not occur.  } The resultant baryon asymmetry can be
estimated as
\begin{equation}
	r \;\simeq\; \left \{
	\begin{array}{ll}
	1.3\times 10^{-4} \displaystyle
	\left ( \frac{1~{\rm MeV}}{m_{3/2}} \right )^{1/5}
	\left ( \frac{T_R}{10^4~{\rm GeV}} \right )
	\left ( \frac{M}{10^{16}~{\rm GeV}} \right )^{12/5} \sin (n \theta)
	&~~~{\rm for}~~~T_R < T_R^c \\
	3.5 \times 10^{-7}\displaystyle
	\left ( \frac{0.1}{\alpha} \right )^{2}
	\left ( \frac{m_{3/2}}{1~{\rm MeV}} \right ) 
	\left ( \frac{10^8~{\rm GeV}}{T_R} \right )
	\left ( \frac{M}{10^{16}~{\rm GeV}} \right )^{3} \sin (n \theta)
	&~~~{\rm for}~~~T_R > T_R^c 
	\end{array} \right. . \label{rGMSB}
\end{equation}
Thus the isocurvature perturbation is calculated as
\begin{equation}
	S \;\simeq\; \left \{
	\begin{array}{ll}
	4\times 10^{-4}r^{11/16} \displaystyle
	\left ( \frac{1~{\rm MeV}}{m_{3/2}} \right )^{1/16}
	\left ( \frac{T_R}{10^4~{\rm GeV}} \right )^{5/16} 
	\left ( \frac{H_{\rm inf}}{10^{14}~{\rm GeV}} \right )^{3/4}\\
	~~~~~\times \cot (n \theta) \sin^{5/16} (n \theta) 
	~~~{\rm for}~~~T_R < T_R^c \\
	2 \times 10^{-4} r^{3/4}\displaystyle
	\left ( \frac{0.1}{\alpha} \right )^{1/2}
	\left ( \frac{m_{3/2}}{1~{\rm MeV}} \right )^{1/4} 
	\left ( \frac{10^8~{\rm GeV}}{T_R} \right )^{1/4}
	\left ( \frac{H_{\rm inf}}{10^{14}~{\rm GeV}} \right )^{3/4}\\
	~~~~~\times \cot (n \theta) \sin^{1/4} (n \theta)
	~~~{\rm for}~~~T_R > T_R^c 
	\end{array} \right. ,  \label{SGMSB}
\end{equation}
if the linear term in Eq.~(\ref{Sb}) dominates.
The non-linearity parameter can be calculated as
\begin{equation}
	f_{\rm NL}^{\rm (iso)} \;\simeq\; \left \{
	\begin{array}{ll}
	-5\times 10^{2}r^{7/4} \displaystyle
	\left ( \frac{1~{\rm MeV}}{m_{3/2}} \right )^{1/4}
	\left ( \frac{T_R}{10^4~{\rm GeV}} \right )^{5/4} 
	\left ( \frac{H_{\rm inf}}{10^{14}~{\rm GeV}} \right )^{3}\\
	~~~~~\times \cot^2 (n \theta) \sin^{5/4} (n \theta) 
	~~~{\rm for}~~~T_R < T_R^c \\
	-1 \times 10^{} r^2\displaystyle
	\left ( \frac{0.1}{\alpha} \right )^{2}
	\left ( \frac{m_{3/2}}{1~{\rm MeV}} \right ) 
	\left ( \frac{10^8~{\rm GeV}}{T_R} \right )
	\left ( \frac{H_{\rm inf}}{10^{14}~{\rm GeV}} \right )^{3}\\
	~~~~~\times \cot^2 (n \theta) \sin^{} (n \theta)
	~~~{\rm for}~~~T_R > T_R^c   \label{fNLGMSB}
	\end{array} \right. .
\end{equation}
If the parameters take appropriate values, $f_{\rm NL}^{\rm (iso)}$ can be large enough with either positive or negative sign.

In GMSB models, there also exist a $Q$-ball solution, but
its properties differ from those of the gravity-mediation type.
The charge of the $Q$-ball is estimated as
\begin{equation}
	Q \;\simeq\; \beta \left ( \frac{\phi_{\rm os}}{M_F} \right )^4
	\times \left \{
	\begin{array}{ll}
		\epsilon ~~~&{\rm for}~~~\epsilon > \epsilon_c \\
		\epsilon_c ~~~&{\rm for}~~~\epsilon < \epsilon_c
	\end{array} \right. ,
\end{equation}
where $\beta \sim 6\times 10^{-4}$ and $\epsilon_c \sim 0.06$~\cite{Kasuya:1999wu} 
and $M_F$ in the denominator should be replaced with 
$T_{\rm os}$ if the oscillation begins due to thermal effects.
If the $Q$-ball energy per unit charge ($M_Q/Q \sim M_F Q^{-1/4}$) 
is smaller than the nucleon mass $\sim 1$~GeV,
$Q$-balls become stable against the decay into nucleons
and contribute to some fraction of the dark matter \cite{Kusenko:1997si}.
Also in that case only the evaporated charges in the high-temperature plasma 
provides the baryon number remaining in the universe \cite{Laine:1998rg,Banerjee:2000mb}. 
In the regime of $\epsilon < \epsilon_c$ the $Q$-ball charge can be calculated as
\begin{equation}
	Q \;\simeq\; \left \{
	\begin{array}{ll}
	5\times 10^{17} \displaystyle
	\left ( \frac{1~{\rm MeV}}{m_{3/2}} \right )^{6/5}
	\left ( \frac{M}{10^{16}~{\rm GeV}} \right )^{12/5}
	&~~~{\rm for}~~~T_R < T_R^c \\
	2 \times 10^{11}\displaystyle
	\left ( \frac{0.1}{\alpha} \right )^{2}
	\left ( \frac{10^8~{\rm GeV}}{T_R} \right )^2
	\left ( \frac{M}{10^{16}~{\rm GeV}} \right )^{3}
	&~~~{\rm for}~~~T_R > T_R^c 
	\end{array} \right. .
	\label{QGMSB}
\end{equation}
One can check that $1~$GeV $< M_Q/Q$ is always met in the parameter
regions of our concern.  Thus $Q$-balls are unstable in the most of the
interesting parameter region and $Q$-ball formation does not affect
following results. (Note that no LSPs are produced by the $Q$-ball decay
since it is kinematically forbidden.)

\begin{figure}[t]
   \includegraphics[width=0.8\linewidth]{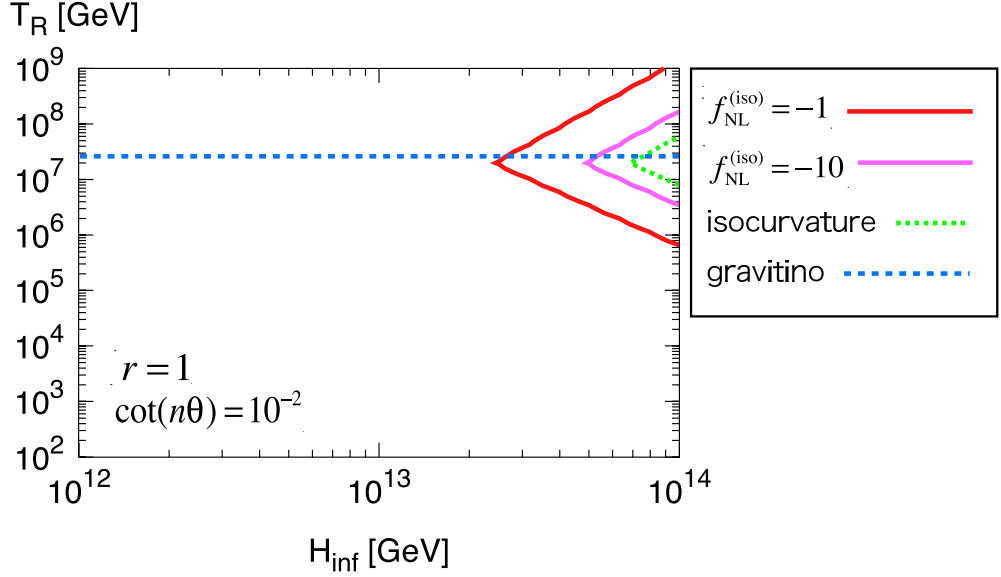}
   \caption{ 
   Same as Fig.~\ref{fig:ADm1TeV}, but for the gauge mediation. 
   We have set $m_{3/2}=10$~GeV, $r=1$ and $\cot (n \theta) =10^{-2}$.
    }
   \label{fig:ADm10}
\end{figure}

Assuming that the AD mechanism creates the total baryon number of the Universe
($r=1$), we obtain constraints on $(H_{\rm inf},T_R)$ plane as shown
in Fig.~\ref{fig:ADm10}. We have set $m_{3/2}=10$~GeV and $\cot (n \theta)
=10^{-2}$.  Meanings of lines are same as Fig.~\ref{fig:ADm1TeV}.  
We can see that a significant amount of non-Gaussianity can be generated,
similar to the gravity-mediation case in the previous subsection.
However, as we decrease the gravitino mass, the allowed region will become
smaller.


We are also interested in the case that the AD mechanism generates
a small fraction of the baryon asymmetry, with both positive and negative signs.
Substituting (\ref{rGMSB}) into Eqs.~(\ref{SGMSB}) and (\ref{fNLGMSB})
yields
\begin{equation}
	S \simeq \left \{
	\begin{array}{ll}
	8\times 10^{-7} \displaystyle
	\left ( \frac{1~{\rm MeV}}{m_{3/2}} \right )^{1/5}
	\left ( \frac{T_R}{10^4~{\rm GeV}} \right )
	\left ( \frac{M}{10^{16}~{\rm GeV}} \right )^{33/20} 
	\left ( \frac{H_{\rm inf}}{10^{14}~{\rm GeV}} \right )^{3/4} \\ 
	~~~~~\times \cot (n \theta) \sin(n\theta) 
	~~~{\rm for}~~~T_R < T_R^c \\
	2 \times 10^{-9}\displaystyle
	\left ( \frac{0.1}{\alpha} \right )^{2}
	\left ( \frac{m_{3/2}}{1~{\rm MeV}} \right ) 
	\left ( \frac{10^8~{\rm GeV}}{T_R} \right )
	\left ( \frac{M}{10^{16}~{\rm GeV}} \right )^{9/4} 
	\left ( \frac{H_{\rm inf}}{10^{14}~{\rm GeV}} \right )^{3/4} \\
	~~~~~\times \cot (n \theta) \sin (n\theta)
	~~~{\rm for}~~~T_R > T_R^c 
	\end{array} \right. .
\end{equation}
and
\begin{equation}
	f_{\rm NL}^{\rm (iso)} \simeq \left \{
	\begin{array}{ll}
	-9\times 10^{-5} \displaystyle
	\left ( \frac{1~{\rm MeV}}{m_{3/2}} \right )^{3/5}
	\left ( \frac{T_R}{10^4~{\rm GeV}} \right )^3
	\left ( \frac{M}{10^{16}~{\rm GeV}} \right )^{21/5} 
	\left ( \frac{H_{\rm inf}}{10^{14}~{\rm GeV}} \right )^{3}\\
	~~~~~\times \cot^2 (n \theta) \sin^3 (n \theta) 
	~~~{\rm for}~~~T_R < T_R^c \\
	-1 \times 10^{-12}\displaystyle
	\left ( \frac{0.1}{\alpha} \right )^{6}
	\left ( \frac{m_{3/2}}{1~{\rm MeV}} \right )^3 
	\left ( \frac{10^8~{\rm GeV}}{T_R} \right )^3
	\left ( \frac{M}{10^{16}~{\rm GeV}} \right )^{6} 
	\left ( \frac{H_{\rm inf}}{10^{14}~{\rm GeV}} \right )^{3}\\
	~~~~~\times \cot^2 (n \theta) \sin^3 (n \theta)
	~~~{\rm for}~~~T_R > T_R^c 
	\end{array} \right. .
\end{equation}

\begin{figure}[t]
   \includegraphics[width=0.8\linewidth]{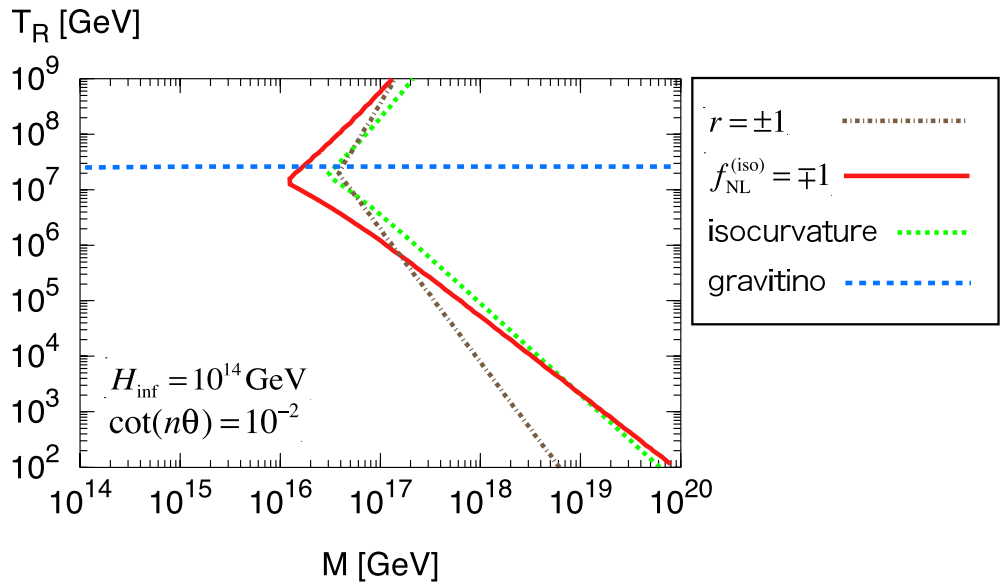}
   \caption{
   Same as Fig.~\ref{fig:ADm1TeVM}, but for the gauge mediation. 
   We have set $m_{3/2}=10$~GeV, $H_{\rm inf}=10^{14}~$GeV and $\cot (n \theta) =10^{-2}$.
      }
   \label{fig:ADm10M}
\end{figure}

The resultant constraints are shown in Fig.~\ref{fig:ADm10M} on $(M,T_R)$ plane. We have set
$m_{3/2}=10~$GeV, $H_{\rm inf}=10^{14}$~GeV and $\cot (n \theta) =10^{-2}$.
We can see that either positive or negative $f_{\rm NL}^{\rm (iso)}$ can be obtained through the AD mechanism
while satisfying the isocurvature constraint.


\section{Discussion and Conclusions} \label{sec:conclusion}

We have focused on the AD mechanism in this paper, but there are other
baryogenesis scenarios that may induce large non-Gaussianity in the
baryon asymmetry. For instance, it is known that the spontaneous
baryogenesis~\cite{Cohen:1988kt} can lead to the baryonic isocurvature
perturbations, since in the original model a non-vanishing chemical
potential is due to a slow-rolling scalar field. In order to estimate
non-Gaussianity, however, one has to specify how the chemical
potential for the baryon number arises. In the case of the spontaneous
baryogenesis using a flat direction of the SSM~\cite{Chiba:2003vp},
the resultant non-Gaussianity has almost the same features as that in
the case of the AD mechanism.  Another example is non-thermal
leptogenesis using the right-handed sneutrino condensate,
$\tilde{N}$~\cite{Murayama:1993em}. Suppose that $\tilde{N}$ is light
and fluctuating around the origin during inflation. Such a set-up may occur without
fine-tuning, because the origin is the symmetry-enhanced point, and
this is exactly what is considered in the ungaussiton scenario
\cite{Suyama:2008nt}.  The $\tilde{N}$ can generate non-Gaussianity in
both the adiabatic and baryonic isocurvature perturbations.  If the
mass of the right-handed sneutrino is heavy, the baryonic isocurvature
perturbations will become more important than the adiabatic one. In
this scenario, the non-Gaussianity becomes large only if the
non-thermal leptogenesis accounts for a tiny fraction of the baryon
asymmetry.

It is also possible to consider non-Gaussianity in other type of
isocurvature perturbations.  For instance, a large lepton asymmetry
may have isocurvature perturbations with some amount of
non-Gaussianity. If they are generated by the AD field with a lepton
number \cite{Kawasaki:2002hq}, non-Gaussianity arises from the
fluctuations of the phase component, as in the case of the AD
mechanism.  However, such isocurvature perturbations will affect the
CMB temperature fluctuations in a different way, and so does the
associated non-Gaussianity. We leave this issue for future work.

So far we have considered up to the three-point correlation functions.
It is straightforward to extend our analysis to the correlation functions
of higher order. In particular, when the linear perturbation is negligible,\footnote{
	To generate large non-Gaussianity while satisfying the constraint on the amplitude of the 
	isocurvature perturbation, the linear term in Eq.~(\ref{eq:S-expansion}) must be suppressed
	(see also Fig.~\ref{fig:contour}.)
}
we  have
a consistency relation between $\fnl$ and $\tau_{\rm NL}$
as in the case of the ungaussiton~\cite{Suyama:2008nt}, the latter
of which is defined as a non-linearity parameter of the four-point
correlation function.

In summary, in this paper we have studied a scenario that non-Gaussian baryonic
isocurvature fluctuation is produced from the AD mechanism in SUSY.
We have found that the AD mechanism can create large non-Gaussianity
without conflicting with the current isocurvature constraint.
We have seen that for $|f_{\rm NL}^{\rm (iso)}|\gtrsim 1$ 
the Hubble parameter during inflation must be larger than
about $10^{12}$~GeV, and that $|f_{\rm NL}^{\rm (iso)}|$ is bounded as $|f_{\rm NL}^{\rm (iso)}|< 60$
to satisfy the isocurvature constraint.
Interestingly, as opposed to many known mechanisms for generating
sizable non-Gaussianity~\footnote{
	It is pointed out that negative $f_{\rm NL}$ can be obtained
        in the curvaton scenario if the curvaton potential deviates
        from quadratic one \cite{Enqvist:2008gk}.
}, the AD field can generate large non-Gaussianity even if it is the
main component for generating the baryon asymmetry. The non-linerity
parameter $f_{\rm NL}^{\rm (iso)}$ is negative if the AD mechanism is responsible
for the total baryon asymmetry of the Universe, while it can be either
positive or negative otherwise.

In our scenario the non-Gaussianity is necessarily accompanied with
some amount of the isocurvature perturbations, therefore, if there are
indeed large non-Gaussianity arising from the isocurvature
perturbations, the future (or even on-going) observations will detect
the isocurvature perturbations.
It would be very interesting if the non-Gaussianity tells us about the origin of the baryon asymmetry,
which is difficult to be probed, otherwise.

\begin{acknowledgements}

K.N. would like to thank the Japan Society for the Promotion of
Science for financial support.  This work is supported by Grant-in-Aid
for Scientific research from the Ministry of Education, Science,
Sports, and Culture (MEXT), Japan, No.14102004 (M.K.)  and also by
World Premier International Research Center InitiativeiWPI
Initiative), MEXT, Japan.

\end{acknowledgements}




\begin{thebibliography}{}



\bibitem{Komatsu:2008hk}
  E.~Komatsu {\it et al.}  [WMAP Collaboration],
  arXiv:0803.0547 [astro-ph].
  
  
\bibitem{Yadav:2007yy}
  A.~P.~S.~Yadav and B.~D.~Wandelt,
  Phys.\ Rev.\ Lett.\  {\bf 100}, 181301 (2008)
  [arXiv:0712.1148 [astro-ph]].
  
  
\bibitem{Lyth:2001nq}
  D.~H.~Lyth and D.~Wands,
  Phys.\ Lett.\  B {\bf 524}, 5 (2002)
  [arXiv:hep-ph/0110002];
  T.~Moroi and T.~Takahashi,
  Phys.\ Lett.\  B {\bf 522}, 215 (2001)
  [Erratum-ibid.\  B {\bf 539}, 303 (2002)]
  [arXiv:hep-ph/0110096];
  K.~Enqvist and M.~S.~Sloth,
  Nucl.\ Phys.\  B {\bf 626}, 395 (2002)
  [arXiv:hep-ph/0109214].
  
  
\bibitem{Lyth:2002my}
  D.~H.~Lyth, C.~Ungarelli and D.~Wands,
  Phys.\ Rev.\  D {\bf 67}, 023503 (2003)
  [arXiv:astro-ph/0208055].
  
  
\bibitem{Suyama:2008nt}
  T.~Suyama and F.~Takahashi,
  arXiv:0804.0425 [astro-ph].
  
  
\bibitem{Linde:1996gt}
  A.~D.~Linde and V.~F.~Mukhanov,
  Phys.\ Rev.\  D {\bf 56}, 535 (1997)
  [arXiv:astro-ph/9610219].
 
\bibitem{Boubekeur:2005fj}
  L.~Boubekeur and D.~H.~Lyth,
  Phys.\ Rev.\  D {\bf 73}, 021301 (2006)
  [arXiv:astro-ph/0504046].
  
  
\bibitem{Kawasaki:2008sn}
  M.~Kawasaki, K.~Nakayama, T.~Sekiguchi, T.~Suyama and F.~Takahashi,
  arXiv:0808.0009 [astro-ph];
  arXiv:0810.0208 [astro-ph].
  
     
\bibitem{Affleck:1984fy}
  I.~Affleck and M.~Dine,
  Nucl.\ Phys.\  B {\bf 249}, 361 (1985).
  
  
\bibitem{Dine:1995kz}
  M.~Dine, L.~Randall and S.~D.~Thomas,
  Nucl.\ Phys.\  B {\bf 458}, 291 (1996)
  [arXiv:hep-ph/9507453].
 
   
\bibitem{Coleman:1985ki}
  S.~R.~Coleman,
  Nucl.\ Phys.\  B {\bf 262}, 263 (1985)
  [Erratum-ibid.\  B {\bf 269}, 744 (1986)].
  
  
\bibitem{Kusenko:1997zq}
  A.~Kusenko,
  Phys.\ Lett.\  B {\bf 405}, 108 (1997)
  [arXiv:hep-ph/9704273];
  Phys.\ Lett.\  B {\bf 404}, 285 (1997)
  [arXiv:hep-th/9704073].
  
  
\bibitem{Dvali:1997qv}
  G.~R.~Dvali, A.~Kusenko and M.~E.~Shaposhnikov,
  Phys.\ Lett.\  B {\bf 417}, 99 (1998)
  [arXiv:hep-ph/9707423].
  
  
\bibitem{Enqvist:1997si}
  K.~Enqvist and J.~McDonald,
  Phys.\ Lett.\  B {\bf 425}, 309 (1998)
  [arXiv:hep-ph/9711514];
  Nucl.\ Phys.\  B {\bf 538}, 321 (1999)
  [arXiv:hep-ph/9803380];
  Nucl.\ Phys.\  B {\bf 570}, 407 (2000)
  [arXiv:hep-ph/9908316];
  K.~Enqvist, A.~Jokinen and J.~McDonald,
  Phys.\ Lett.\  B {\bf 483}, 191 (2000)
  [arXiv:hep-ph/0004050].


\bibitem{Linde:1985gh}
  A.~D.~Linde,
  Phys.\ Lett.\  B {\bf 160}, 243 (1985).
  
  
\bibitem{Enqvist:1998pf}
  K.~Enqvist and J.~McDonald,
  Phys.\ Rev.\ Lett.\  {\bf 83}, 2510 (1999)
  [arXiv:hep-ph/9811412];
  Phys.\ Rev.\  D {\bf 62}, 043502 (2000)
  [arXiv:hep-ph/9912478];
  M.~Kawasaki and F.~Takahashi,
  Phys.\ Lett.\  B {\bf 516}, 388 (2001)
  [arXiv:hep-ph/0105134].


\bibitem{Kasuya:2008xp}
  S.~Kasuya, M.~Kawasaki and F.~Takahashi,
  arXiv:0805.4245 [hep-ph].
    
    
\bibitem{Babich:2004gb}
  D.~Babich, P.~Creminelli and M.~Zaldarriaga,
  JCAP {\bf 0408}, 009 (2004)
  [arXiv:astro-ph/0405356].
  
  
\bibitem{Lyth:1991ub}
  D.~H.~Lyth,
  Phys.\ Rev.\  D {\bf 45}, 3394 (1992).
  
      
\bibitem{Lyth:2007jh}
  D.~H.~Lyth,
  JCAP {\bf 0712}, 016 (2007)
  [arXiv:0707.0361 [astro-ph]].
  
    
\bibitem{Gherghetta:1995dv}
  T.~Gherghetta, C.~F.~Kolda and S.~P.~Martin,
  Nucl.\ Phys.\  B {\bf 468}, 37 (1996)
  [arXiv:hep-ph/9510370].
  
    
\bibitem{Allahverdi:2000zd}
  R.~Allahverdi, B.~A.~Campbell and J.~R.~Ellis,
  Nucl.\ Phys.\  B {\bf 579}, 355 (2000)
  [arXiv:hep-ph/0001122].


\bibitem{Anisimov:2000wx}
  A.~Anisimov and M.~Dine,
  Nucl.\ Phys.\  B {\bf 619}, 729 (2001)
  [arXiv:hep-ph/0008058].
  
  
\bibitem{Kasuya:2003va}
  S.~Kasuya, M.~Kawasaki and F.~Takahashi,
  Phys.\ Lett.\  B {\bf 578}, 259 (2004)
  [arXiv:hep-ph/0305134].
  
  
\bibitem{Kasuya:2003yr}
  S.~Kasuya, M.~Kawasaki and F.~Takahashi,
  Phys.\ Rev.\  D {\bf 68}, 023501 (2003)
  [arXiv:hep-ph/0302154].
  
  
\bibitem{Fujii:2001zr}
  M.~Fujii, K.~Hamaguchi and T.~Yanagida,
  Phys.\ Rev.\  D {\bf 63}, 123513 (2001)
  [arXiv:hep-ph/0102187];
  M.~Kawasaki and K.~Nakayama,
  JCAP {\bf 0702}, 002 (2007)
  [arXiv:hep-ph/0611320].


\bibitem{Riotto:2008gs}
  A.~Riotto and F.~Riva,
  arXiv:0806.3382 [hep-ph].
  
                   
\bibitem{Moroi:1993mb}
  T.~Moroi, H.~Murayama and M.~Yamaguchi,
  Phys.\ Lett.\  B {\bf 303}, 289 (1993).
  
  
\bibitem{Kawasaki:2004yh}
  M.~Kawasaki, K.~Kohri and T.~Moroi,
  Phys.\ Lett.\  B {\bf 625}, 7 (2005)
  [arXiv:astro-ph/0402490];
  Phys.\ Rev.\  D {\bf 71}, 083502 (2005)
  [arXiv:astro-ph/0408426];
  M.~Kawasaki, K.~Kohri, T.~Moroi and A.~Yotsuyanagi,
  arXiv:0804.3745 [hep-ph].
  
  
\bibitem{Kawasaki:2005xc}
  M.~Kawasaki, K.~Konya and F.~Takahashi,
  Phys.\ Lett.\  B {\bf 619}, 233 (2005)
  [arXiv:hep-ph/0504105].
  
  
\bibitem{Kasuya:1999wu}
  S.~Kasuya and M.~Kawasaki,
  Phys.\ Rev.\  D {\bf 61}, 041301 (2000)
  [arXiv:hep-ph/9909509];
  Phys.\ Rev.\  D {\bf 64}, 123515 (2001)
  [arXiv:hep-ph/0106119].
  
  
\bibitem{Fujii:2001xp}
  M.~Fujii and K.~Hamaguchi,
  Phys.\ Lett.\  B {\bf 525}, 143 (2002)
  [arXiv:hep-ph/0110072];
  Phys.\ Rev.\  D {\bf 66}, 083501 (2002)
  [arXiv:hep-ph/0205044];
  M.~Fujii and T.~Yanagida,
  Phys.\ Lett.\  B {\bf 542}, 80 (2002)
  [arXiv:hep-ph/0206066].
  
  
\bibitem{Seto:2005pj}
  O.~Seto,
  Phys.\ Rev.\  D {\bf 73}, 043509 (2006)
  [arXiv:hep-ph/0512071].
  
    
\bibitem{Kawasaki:2007yy}
  M.~Kawasaki and K.~Nakayama,
  Phys.\ Rev.\  D {\bf 76}, 043502 (2007)
  [arXiv:0705.0079 [hep-ph]].
 
\bibitem{Hamaguchi:2003dc}
  K.~Hamaguchi, M.~Kawasaki, T.~Moroi and F.~Takahashi,
  Phys.\ Rev.\  D {\bf 69}, 063504 (2004)
  [arXiv:hep-ph/0308174].
 
  
  
\bibitem{Giudice:1998bp}
  For a review, see G.~F.~Giudice and R.~Rattazzi,
  Phys.\ Rept.\  {\bf 322}, 419 (1999)
  [arXiv:hep-ph/9801271].
  
  
\bibitem{de Gouvea:1997tn}
  A.~de Gouvea, T.~Moroi and H.~Murayama,
  Phys.\ Rev.\  D {\bf 56}, 1281 (1997)
  [arXiv:hep-ph/9701244].
  
      
\bibitem{Kusenko:1997si}
  A.~Kusenko and M.~E.~Shaposhnikov,
  Phys.\ Lett.\  B {\bf 418}, 46 (1998)
  [arXiv:hep-ph/9709492].
  
  
\bibitem{Laine:1998rg}
  M.~Laine and M.~E.~Shaposhnikov,
  Nucl.\ Phys.\  B {\bf 532}, 376 (1998)
  [arXiv:hep-ph/9804237].
  
  
\bibitem{Banerjee:2000mb}
  R.~Banerjee and K.~Jedamzik,
  Phys.\ Lett.\  B {\bf 484}, 278 (2000)
  [arXiv:hep-ph/0005031].
    
    
\bibitem{Cohen:1988kt}
  A.~G.~Cohen and D.~B.~Kaplan,
  Nucl.\ Phys.\  B {\bf 308}, 913 (1988);
  A.~G.~Cohen, D.~B.~Kaplan and A.~E.~Nelson,
  Phys.\ Lett.\  B {\bf 263}, 86 (1991).
    
    
\bibitem{Chiba:2003vp}
  T.~Chiba, F.~Takahashi and M.~Yamaguchi,
  Phys.\ Rev.\ Lett.\  {\bf 92}, 011301 (2004)
  [arXiv:hep-ph/0304102];
  F.~Takahashi and M.~Yamaguchi,
  Phys.\ Rev.\  D {\bf 69}, 083506 (2004)
  [arXiv:hep-ph/0308173].
    
\bibitem{Kawasaki:2002hq}
  M.~Kawasaki, F.~Takahashi and M.~Yamaguchi,
  Phys.\ Rev.\  D {\bf 66}, 043516 (2002)
  [arXiv:hep-ph/0205101].
   
      
\bibitem{Murayama:1993em}
  H.~Murayama and T.~Yanagida,
  Phys.\ Lett.\  B {\bf 322}, 349 (1994)
  [arXiv:hep-ph/9310297];
  K.~Hamaguchi, H.~Murayama and T.~Yanagida,
  Phys.\ Rev.\  D {\bf 65}, 043512 (2002)
  [arXiv:hep-ph/0109030].

 
\bibitem{Enqvist:2008gk}
  K.~Enqvist and T.~Takahashi,
  arXiv:0807.3069 [astro-ph];
  Q.~G.~Huang and Y.~Wang,
  arXiv:0808.1168 [hep-th].
  



\end{thebibliography}
\end{document}